\documentclass[showkeys,prb,reprint]{revtex4-1}
\usepackage{hyperref}
\usepackage{graphicx}
\usepackage{float}

\begin{document}

%-------------------------------------------------------------------------------------------------

\title{A short note on the SIMS characterisation of ZnO/Si(100) interface}

\author{Kumar Kumarappan}
\email[]{kumar.kumarappan2@mail.dcu.ie}
\affiliation{School of Physical Sciences and National Centre for Sensor Research, Dublin City University, Dublin-9, Ireland.}

%-------------------------------------------------------------------------------------------------

\begin{abstract}

The ZnO films, 25 nm thick were deposited by e-beam evaporation of ZnO (99.9$\%$) pellets onto native oxide covered silicon (100) substrates. Details of the interface chemical composition and the chemical depth profile have been deduced as follows: Adsorbed OH surface/OH: ZnO Interface/Pure ZnO thin film/ZnO:H: Si Interface/H:SiO2/SiO2/Si(100). The physical profile of the crater formed by the SIMS depth profiling was measured with a Tencor P-2 profilometer as depth profiling showed uniform ion milling and good consistency with film thickness (25 nm) measured by SIMS depth profile (25.4 nm).

\end{abstract}

\keywords{Secondary Ion Mass Spectrometery; Semiconductor Surfaces; Interface Analysis; ZnO; Silicon Surface}

\maketitle

%-------------------------------------------------------------------------------------------------
\section{Introduction}
Dynamic-Secondary Ion Mass Spectroscopy (D-SIMS) has been used to acquire a chemical depth profile of e-beam deposited ZnO thin films. SIMS has been used previously to detect elemental impurities in ZnO~\cite{Sakaguchi, McCluskey} and to detect the presence of residual gold in ZnO nanorod grown by the VPT process~\cite{Morris}. In this study SIMS is used to identify the surface contamination on ZnO surface, obtain a depth profile of the thin film chemical composition and study the interface composition with a silicon substrate in order to derive a depth profile model.
\section{Experimental}
The ZnO films, 25 nm thick were deposited by e-beam evaporation of ZnO (99.9$\%$) pellets purchased from ABCR GmbH $\&$ Co. KG, Germany onto native oxide covered silicon (100) substrates. The distance from the e-beam source to both the substrate surface and the thin film quartz crystal monitor was $\approx$15 cm. The base pressure in the vacuum system before deposition was 5 x 10$^{-7}$ mbar, while during evaporation the pressure typically rose to 7 x $10^{-5}$ mbar. For e-beam settings of 5 kV and beam current of 26.3 mA, the measured deposition rate was 0.01 nm/second. Both positive and negative SIMS depth profiles were acquired using an incident neutral 5~kV Ar beam. The sample current was measured to be 40 nA, which resulted in an etching rate of approximately 0.85 nm/minute. The SIMS spectra were acquired at a background chamber pressure of 1x10$^{-7}$mbar argon. The physical profile of the crater formed by the SIMS depth profiling was measured with a Tencor P-2 profilometer.
\section{Result and Discussion}
\begin{figure}
	\centering
		\includegraphics [scale=0.5]{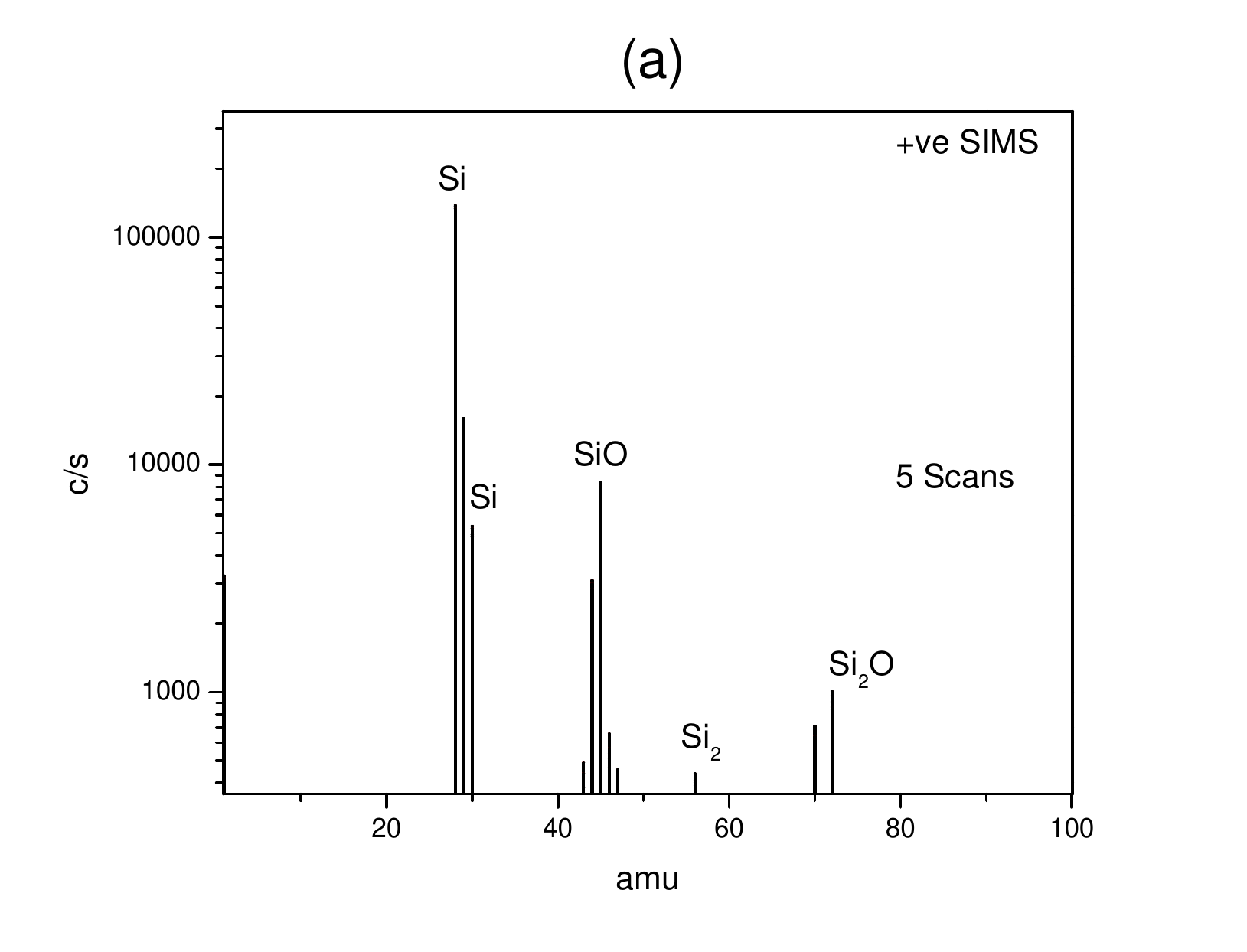}
	\caption{Mass spectrum for Silicon substrate}
	\label{fig:Graph08}
			 	\end{figure}

Dynamic SIMS profiles were acquired for ZnO thin films with an estimated thickness of 25 nm grown on silicon substrates. Mass spectra were initially acquired from different regions of silicon surface with and without the ZnO film. The SIMS bar analysis as shown in figure \ref{fig:Graph08} reflects the profile of the native oxide covered silicon surface with the detection of the isotopes of silicon ($^{28}Si^{+}$, $^{29}Si^{+}$, $^{30}Si^{+}$) and oxides of silicon SiO(m/z = 44), Si$_{2}O$ (m/z = 72). This spectrum is consistent with the abundance of Si isotopes and oxide species of a standard reference of silicon oxide~\cite{Magee} and was measured to ensure correct operation of the instrument. 

\begin{figure}
	\centering
		\includegraphics [scale=0.5]{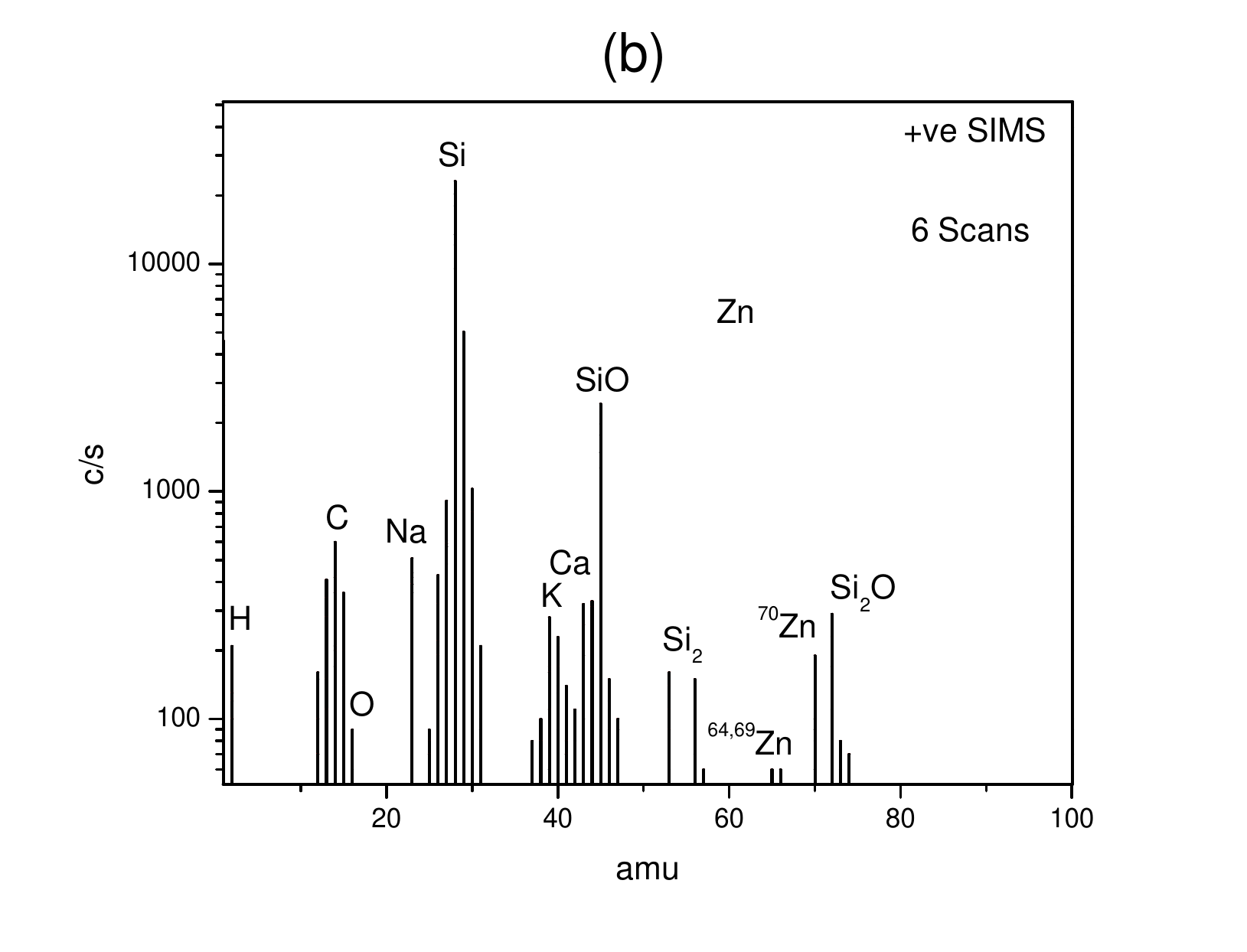}
	\caption{Mass spectrum of edge of the sample}
	\label{fig:Graph09}
		 	\end{figure}
	 	
\begin{figure}
	\centering
		\includegraphics [scale=0.5]{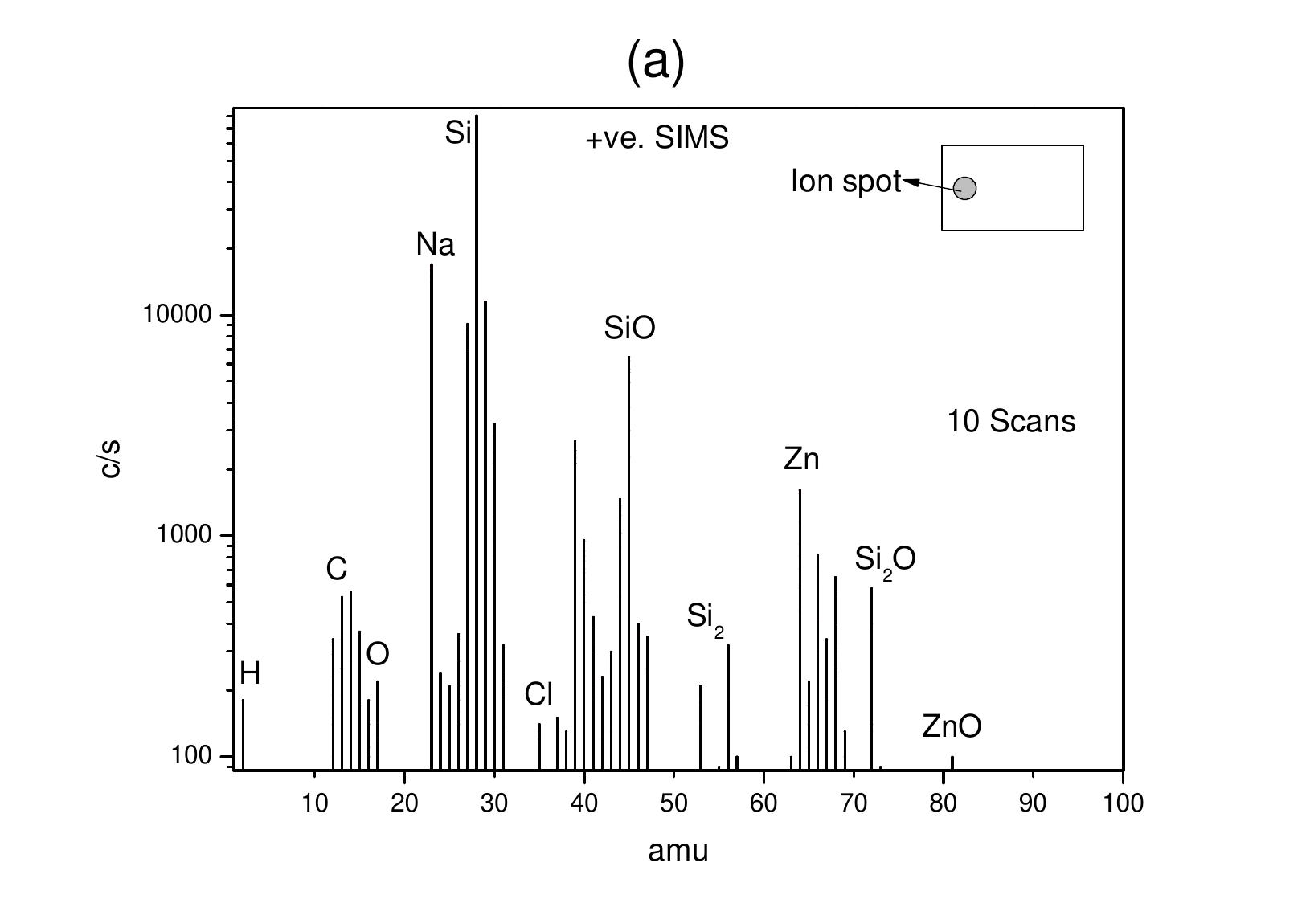}
	\caption{Mass spectra of (edge of film)ZnO thin film}
	\label{fig:Graph10}
		 	\end{figure}

\begin{figure}
	\centering
		\includegraphics [scale=0.5]{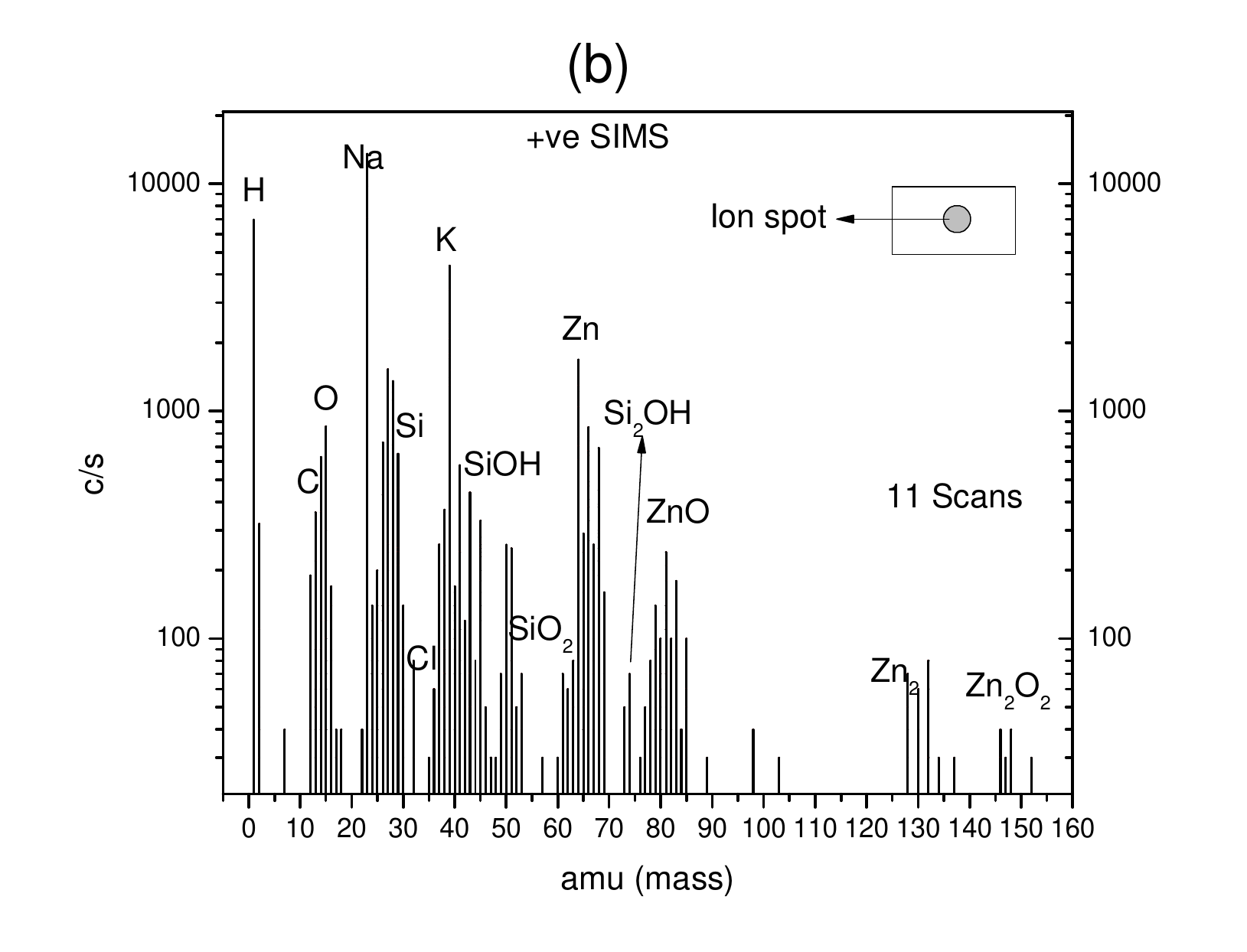}
	\caption{abundance region (Centre of sample, uniform film) of ZnO film}
	\label{fig:Graph11}
			\end{figure}

\begin{figure}
	\centering
		\includegraphics[scale=0.5]{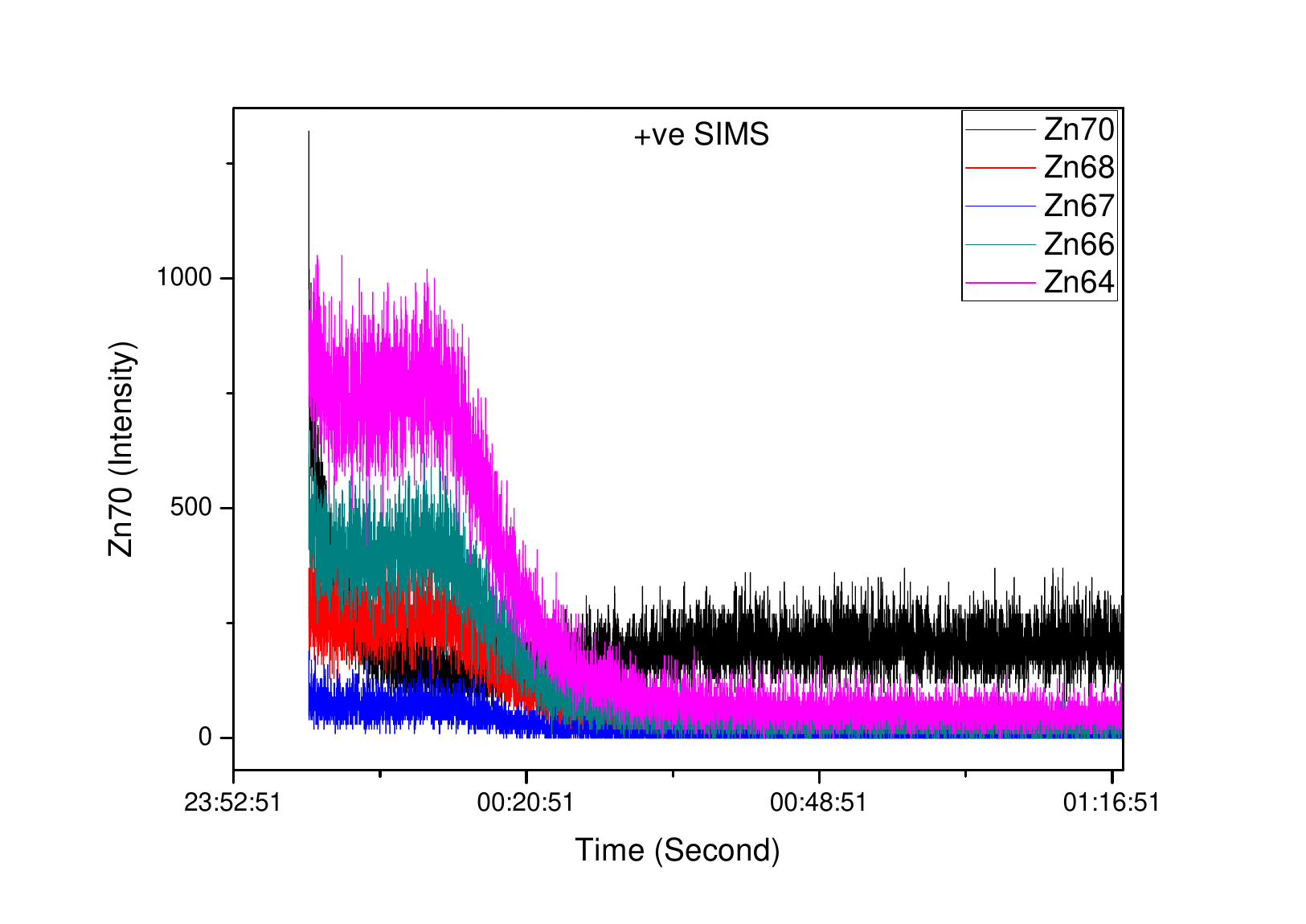}
		\caption{Isotope analysis of Zinc element for detection of abundance}
		\label{fig:Graph12}
\end{figure}

Figure \ref{fig:Graph09} shows the mass profile when the SIMS argon beam is incident on the edge of the sample as it displays the presence of both ZnO related species with isotopes of Zn (mass = 64, 67 $\&$ 70) and O (m/z = 16) as well as less intensity substrate related silicon Si(m/z = 28), $Si_{2}$(m/z = 56) its oxides $^{28}$Si$^{+ 16}$O$^{-}$, $^{44}$Si$_{2}^{+32}$O$^{-}$. The absence of a ZnO(m/z = 80) peak maybe evidence of the poor growth of ZnO in edge of sample.  Common contaminants widely detected in many SIMS profiles of a wide range of materials such as  $^{1}H^{+}$,   $^{12}C^{+}$, $^{13}C^{+}$, $^{23}Na^{+}$, $^{39}K^{+}$, $^{40}Ca^{+}$ are clearly observed in the film composition. When the primary $Ar^{+}$ ion is moved fully onto the ZnO film, high intensity mass spectra for Zn ion and zinc oxide compounds are clearly observed in figure \ref{fig:Graph10}. The commonly detected natural contaminant of chlorine in ZnO material is present in this spectrum as its isotopes $^{35}Cl^{-}$ and $^{37}Cl^{-}$ are seen. The SIMS-intensity distribution for pure ZnO target matched well with the natural abundance of Zn and O combinations previously reported by Meyer et al~\cite{Meyer}. Again the substrate silicon and other basic contaminants $^{1}H^{+}$, $^{12}C^{+}$,$^{13}C^{+}$, $^{23}Na^{+}$, $^{39}K^{+}$, $^{40}Ca^{+}$, were found. The mass spectrum shown in figure \ref{fig:Graph11} displays very high intense peaks for both Zn isotopes $^{64}Zn^{+}$, $^{66}Zn^{+}$, $^{67}Zn^{+}$, $^{68}Zn^{+}$, $^{70}Zn^{+}$ and all mass region of zinc oxide $^{64}Zn^{+16}O^{-}$, $^{66}Zn^{+16}O^{-}$, $^{67}Zn^{+16}O^{-}$, $^{68}Zn^{+16}O^{-}$, $^{70}Zn^{+16}O^{-}$, even double ZnO species such as $^{128}Zn^{+32}O^{-}$, $^{132}Zn^{+32}O^{-}$, $^{136}Zn^{+32}O^{-}$, $^{140}Zn^{+32}O^{-}$ are detected. These results are consistent with reported SIMS spectra for ZnO~\cite{Sumiya}. A depth profile analysis of zinc isotopes was performed to detect the high abundance isotopes in the ZnO thin films and as shown in figure \ref{fig:Graph12}, the $^{64}Zn^{+}$ isotope which had the highest abundance was subsequently used for depth profile analysis.

\begin{figure}
	\centering
		\includegraphics[scale=0.5]{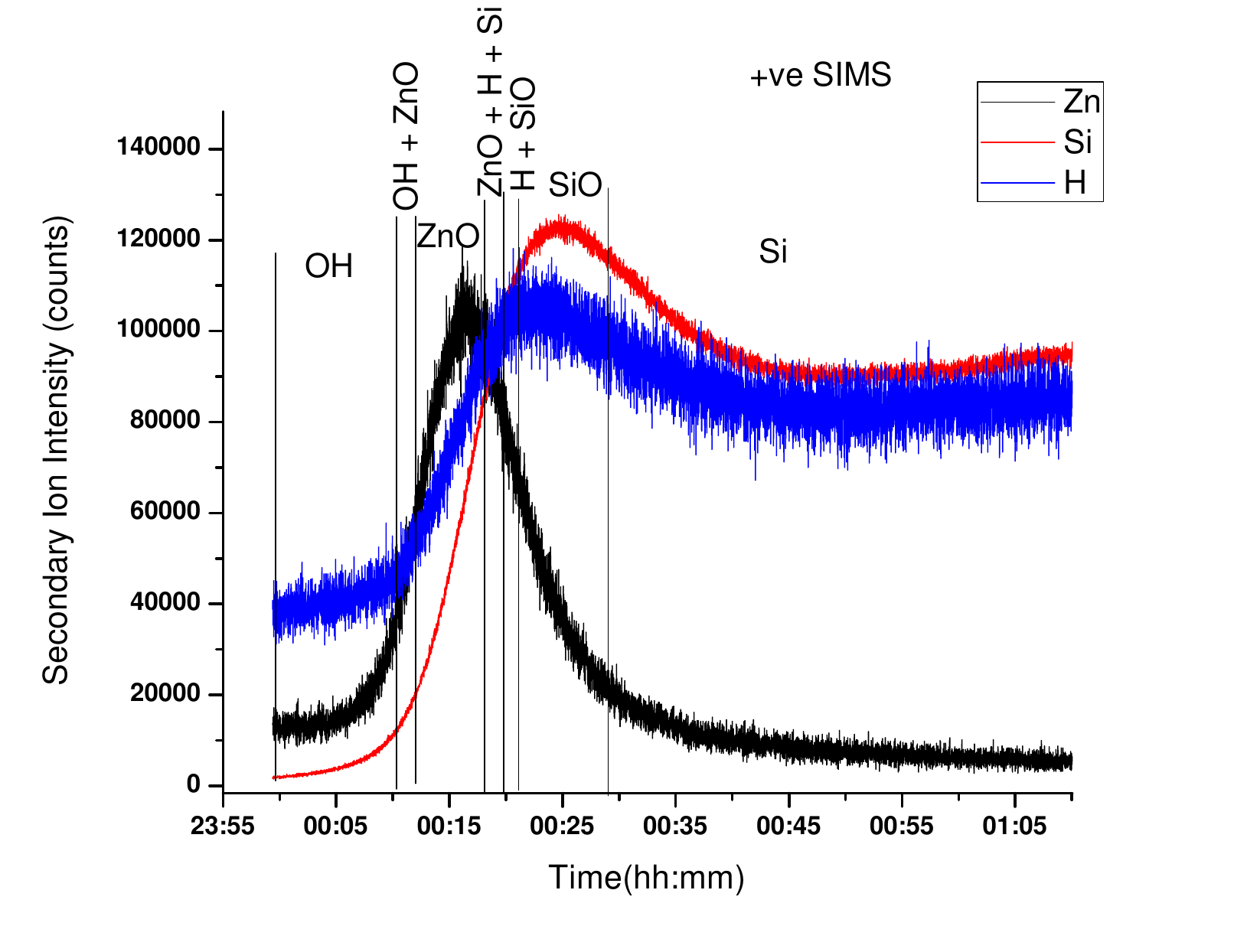}
	\caption{Depth profile of H, Zn $\&$ Si in ZnO(25 nm) grown on Si(100) [where H $\&$ ZnO signal were multiplied by 20}
	\label{fig:Graph13}
	\end{figure}

\begin{figure}
	\centering
		\includegraphics[width=0.45\textwidth]{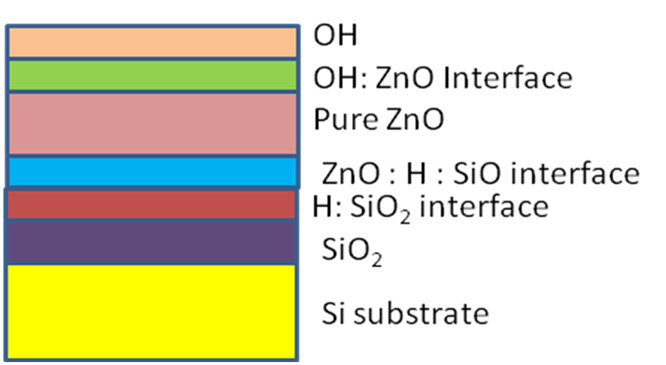}
	\caption{SIMS depth profile model of ZnO(25 nm)/ Si(100) thin film}
	\label{fig:Graph14}
	\end{figure}

Depth profiles were undertaking by monitoring the intensity of $^{2}H^{+}$ for adsorbed hydroxide(OH) on ZnO surface, $^{64}Zn^{+}$ abundance isotope for ZnO and $^{28}Si^{+}$ abundance isotope for silicon substrate. The bombarding was carried out at a base pressure of 1 x 10$^{-7}$ mbar with constant argon flow. The depth profile rate of 0.85 nm/min was estimated from a number of different experimental runs on identically prepared films where the depth of the generated crater was measured by profilometry. The depth profile through the ZnO film shown in figure \ref{fig:Graph13} allows a chemical composition model to be developed. The increase in the silicon signal intensity at the interface between the ZnO film and the silicon substrate is indicative of interface oxidation which reflects the presence of the native oxide on the surface. The pure silicon substrate is seen after bombarding the sample for 43 minutes (36.5 nm). The depth profile model of this ZnO thin film is shown in figure \ref{fig:Graph14}. The surface topography of the SIMS depth profiled ZnO thin films were characterised by profilometer in order to obtain the etch rate. The SIMS generated crater is circular in shape with a diameter of 1.5 to 2 mm and has uniform depth which was measured to determine the etch rate. 

\section{Conclusion}
SIMS analysis of e-beam deposited ZnO thin films on native oxide covered silicon surfaces has been undertaken. Details of the surface chemical composition and the chemical depth profile have been deduced.  The surface topography of the SIMS craters generated during the depth profiling showed uniform ion milling and good consistency with film thickness (25 nm) measured by SIMS depth profile (25.4 nm).

%--------------------------------------------------------

\end{document}